\documentclass{ifacconf}

\newtheorem{mydef}{Definition}
\newtheorem{mythm}{Theorem}
\newtheorem{myprob}{Problem}

\newtheorem{mypro}{Proposition}

\newtheorem{remark}{Remark}

\usepackage{amsfonts}
\DeclareMathSymbol{\shortminus}{\mathbin}{AMSa}{"39}

\usepackage{graphicx}      
\usepackage{natbib}     
\usepackage{multirow}
\usepackage{makecell}
\usepackage{booktabs}
\usepackage{array}
\usepackage{graphicx}\usepackage{url}
\usepackage{subfigure} 
\usepackage{float}
\usepackage{amsmath}
\usepackage{amsfonts}
\usepackage{epsfig}
\usepackage{graphicx}
\usepackage{arydshln}
\usepackage{algpseudocode}
\usepackage{tikz}
\usetikzlibrary {arrows.meta}
\usepackage{verbatim}
\usepackage{subfigure}
\usepackage{enumerate}
\usepackage{rotating}
\usepackage{color}
\usepackage{caption}
\usepackage{flushend}
\usepackage{mathrsfs}
\usepackage{amssymb}
\usepackage[ruled,vlined,linesnumbered]{algorithm2e}
\usepackage{float}
\usepackage{soul}
\usepackage{fancyhdr}
\begin{document}
\begin{frontmatter}

\title{You Don't Know When I Will Arrive: Unpredictable Controller Synthesis for Temporal Logic Tasks  \thanksref{footnoteinfo}} 

\thanks[footnoteinfo]{This work was supported by the National Natural Science Foundation of China (62061136004, 61803259, 61833012). 
This work was also supported in part by NSF CCRI 1925587 and DARPA FA8750-20-C-0542 (Systemic Generative Engineering). The views, opinions, and/or findings expressed are those of the author(s) and should not be interpreted as representing the official views or policies of the Department of Defense or the U.S. Government.
The first two authors contribute equally.}

\author[First]{Yu Chen} 
\author[Second]{Shuo Yang} 
\author[Second]{Rahul Mangharam} 
\author[First]{Xiang Yin}

\address[First]{Department of Automation, Shanghai Jiao Tong
University, Shanghai 200240, China. {e-mail: \{yuchen26,yinxiang\}@sjtu.edu.cn})}
\address[Second]{Department of Electrical and Systems Engineering, University of Pennsylvania, Philadelphia, PA 19104, USA. (e-mail: \{yangs1, rahulm\}@seas.upenn.edu)} 

\begin{abstract}                
In this paper, we investigate the problem of synthesizing  controllers  for temporal logic specifications under security constraint. We assume that there exists a passive intruder (eavesdropper) that can partially observe the behavior of the system. 
For the purpose of security, we require that  the system's behaviors are unpredictable in the sense that the intruder cannot determine for sure that the system will exactly accomplish the task in $K$ steps ahead. This problem is particularly challenging since future information is involved in the synthesis process. We propose a novel information structure that predicts the effect of control in the future. A sound and complete algorithm is developed to synthesize a controller which ensures both task completion and security guarantee. The proposed approach is illustrated by a case study of robot task planning.
\end{abstract}

\begin{keyword}
Information-Flow Security, Robot Path Planning, Temporal Logic Tasks.
\end{keyword}

\end{frontmatter}

\section{Introduction}

With the development of information technologies, autonomous systems such as robotics are required to accomplish complex tasks in dynamic environments. Temporal logics, such as linear temporal logic (LTL), have drawn considerable attention in the control community in the recent years due to their high expressibility in  describing complex tasks. Using automata-theoretic approaches, high-level controllers can be effectively synthesized  to achieve complex temporal logic tasks in dynamic environment~\cite{kress2018synthesis}. 

When executing tasks, the system usually need to communicate with other components such as clouds or central stations to exchange information. However, the communication process may leak critical information to the outside world. For example, there might exist an eavesdropper  ``listening'' the communication between a robot and a central station. Then, the intruder may use this information to  infer whether or not the robot is performing some secret task. 
Therefore, in addition to the correctness requirement, 
security issue has also been becoming increasingly important consideration in controller synthesis \cite{liu2022secure}.

\emph{Our Contributions: }
In this paper, we formulate and solve a new security-aware controller synthesis problem for co-safe linear temporal logic (scLTL) specifications. We model the system as a non-deterministic transition system. The correctness requirement of the controller is to ensure the satisfaction of a given scLTL task. Furthermore, we assume that there is an eavesdropper/intruder  monitoring the system’s behaviors via an output function.  For the purpose of security, we further require that the behaviors of system are \emph{unpredicable} in the sense that  the intruder can never determine for sure that the system will finish the task at some specific future instant. 

To capture this information-flow security constraint formally, we propose a new security requirement called \emph{$K$-step unpredictability}. Specifically, this notion requires that, for any path in controlled system, there exists another observation-equivalent path that will not accomplish the task in exactly $K$ steps. The difficulty in enforcing such a security requirement is that the synthesis process is ``non-causal''. Specifically, to determine  whether or not the intruder can successfully predict when the task will be completed, the controller should know not only its strategy in the past, but also the strategy to be applied in the future, which is unknown and to be synthesized. 

To resolve this future dependency issue, we propose a novel approach by ``borrowing'' information in the future. The general idea is to make  a \emph{prediction} at each instant regarding in how many number of steps, the scLTL task will be accomplished. Then, we build an information structure in which all predictions are correct in the sense that they are consistent with what actually happens in the future. A backtracking-based algorithm is then proposed to synthesize a controller that achieves both the scLTL task and the unpredictable security requirement.  

\emph{Related Work: }
There has been an increasing interest in high-level controller synthesis for robotic specifications; see, e.g.,
\cite{smith2011optimal,guo2015multi,kloetzer2020path,kantaros2018sampling,cai2020learning,shi2022path}. 
However, these works only focus on the correctness of system without considering the security requirement.
In the context of supervisory control of discrete event systems, algorithms have been developed for enforcing the notion of opacity \cite{yin2016uniform,tong2018current}. 
In the context of security-aware controller synthesis, our work is mostly related to \cite{wang2020hyperproperties,yang2020secure,xie2021secure}, where both LTL specifications and security constraints are enforced.  However, the security constraints in these work require that the intruder can never infer that the robot started from a secret location either initially or at some specific instant. Such security requirements are only related to the past behaviors of the system, while we consider the unpredictability of the \emph{future behavior} of the system. 

\section{Motivating Example}\label{sec:motivating-example}
\begin{figure}[!htp]
\begin{minipage}{0.23\textwidth}
  \centering
  \includegraphics[height=3.3cm]{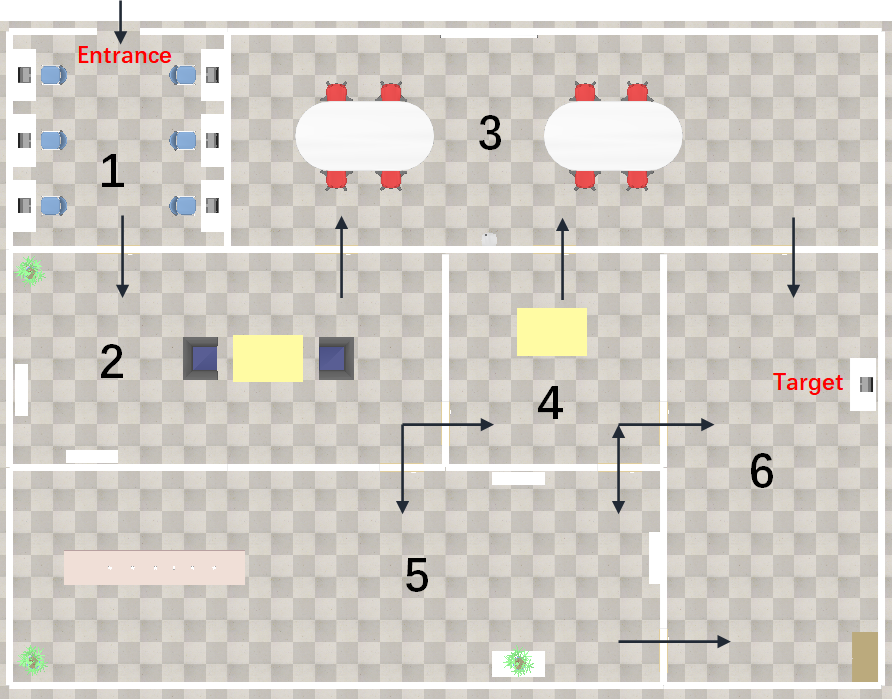}
  \caption{Robot workspace.}
  \label{fig:work1}
\end{minipage}\hfill
\begin{minipage}{0.23\textwidth}
  \centering
  \includegraphics[height=3.3cm]{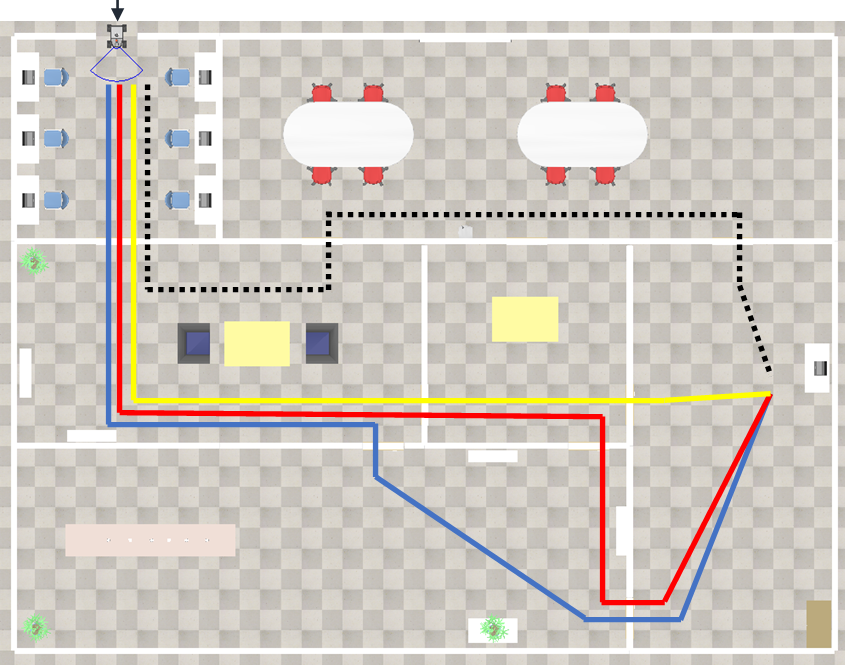}
  \caption{Different paths.}
  \label{fig:tra1}
\end{minipage}
\end{figure}

Before we formulate the problem, we first consider a motivating example. We consider a robot  moving in a workspace with six regions  as shown in Fig.~\ref{fig:work1}. The robot is initialized at Region $1$. Black arrows in the figure represent the permitted moves of the robot from one region to another.  Note that, from Region $2$, the robot cannot move to  Regions $4$ or $5$ for sure since two doors are too closed, i.e., these two transitions are non-deterministic. Similar situation happens when the robot tries to move from from Region $4$ to Region $5$ or $6$.



The objective of the robot is to first reach Region $2$ to collect a critical message and then move to Region $6$ to broadcast it. Assume that there exists an intruder knowing the control policy and the current location of robot. The intruder intends to hack into the broadcasting channel to get the message. However, it only has one chance to hack the channel. Therefore, the intruder needs to predict the precise time instant when the robot will finish the task, i.e., reaching Region $6$ for the first time. 

If the control policy of the robot is to first go to Region $2$ and then Regions $3$ and $6$, 
then there is no non-determinism during the execution. The unique path generated by this policy is shown as the black dashed line in Fig.~\ref{fig:tra1}. Clearly, once robot has started from Region $1$, the intruder knows for sure that the robot will be in Region $6$ after $3$ steps. 
Thus, this control policy is not  ``secure''.
However,  the robot can choose  to go to Region $4$ or $5$ from Region 2, and then go to Region $5$ or $6$ when in Region $4$, and finally go to Region $6$ when in Region $5$.
Under this control policy, robot will generate three possible paths  which are shown as the colored lines  in Fig.~\ref{fig:tra1}. Under this policy, when the robot has started from Region 1, after $3$ steps, it can be in either Region $5$ or $6$.  Therefore, intruder cannot successfully predict that robot will reach Region $6$ in $3$ steps for the first time.

\section{Preliminary}
  
\subsection{System Model}
Let $A$ be a finite set of symbols. 
We denote by $A^{*}$  and $A^\omega$ the sets of all finite and infinite sequences over $A$, respectively, and  
$\epsilon\in A^*$ is the empty sequence.
For any finite sequence $s \!=\! a_{1} \dots a_{n}\!\in\! A^*$, we denote by $|s|\!=\!n$ and $\textsf{last}(s)\!=\!a_n$ its length and its last element, respectively. 

The mobility of the robot in a workspace is modelled as a (non-deterministic) transition system  
\[
G = ( X, x_{0}, U, \rightarrow , \mathcal{AP}, L),
\]
where  
$X$ is the set of \emph{states} representing the locations of the robot;  
$x_{0} \in X$ is the \emph{initial state} of the robot;  
$U$ is the set of \emph{control inputs} or \emph{actions} taken by the robot at each instant;
$\rightarrow \subseteq X \times U \times X $ is the \emph{transition relation}, where $(x,u,x')\in\rightarrow $ (also denoted by $x\xrightarrow{u}x'$) means that the robot can possibly move from state $x$ to state $x'$ under action $u$;
$\mathcal{AP}$ is the set of \emph{atomic propositions} representing some basic properties of our interest; and 
$L : X \to 2^{\mathcal{AP}}$ is the \emph{labeling function}  assigning each state $x$ a set of atomic propositions  that hold at state $x$.

Given an input sequence $u_{1} \dots u_{n} \in U^{*}$, it may induce a finite state \emph{run}
$x_{0} \xrightarrow{u_{1}} x_{1} \xrightarrow{u_{2}} \cdots \xrightarrow{u_{n}} x_{n}$ 
such that  $x_{i} \xrightarrow{u_{i+1}} x_{i+1},\forall i=0,\dots,n-1$; 
state sequence $x_{0}\dots x_{1} \in X^{*}$ is called a finite \emph{path} of system $G$. 
Note that  the run or path induced by an input sequence may not be unique when the system is non-deterministic. 
We denote by $\textsf{Path}(G)$ the set of all finite paths generated by $G$ starting from the initial state $x_{0}$. The \emph{trace} of a  path $\tau=x_{0} \dots x_{n} \in \textsf{Path}(G)$ is defined by $\textsf{trace}(\tau)=L(x_{0})\dots L(x_{n})$, and we denote by  $\textsf{Trace}(G)=\{ \textsf{trace}(\tau): \tau\in \textsf{Path}(G) \}$ the set of all traces in $G$. System $G$ is  called  \emph{live} if for any $x \in X$, there exist $u \in U$ and $x' \in X$ such that $(x,u,x') \in \rightarrow$. We assume that system $G$ is live.
For each state $x\in X$, we define $U(x)=\{ u\in U: \exists x'\in X\text{ s.t. }x\xrightarrow{u}x'\}$ as the set of active inputs at $x$. 
For   a set of states $q\subseteq X$ and    a control input $u\in U$,  
we define the successor states of $q$ under $u$  by
\begin{equation}
    \mathbf{NX}_{u}(q)=\{x'\in X:  \exists x\in q\text{ s.t. }x\xrightarrow{u}x' \}.
\end{equation}

\subsection{Linear Temporal Logic Specifications} 
The task of the robot is described by a co-safe  Linear Temporal Logic (scLTL) formula $\varphi$ over atomic propositions $\mathcal{AP}$. 
Formally, the syntax of scLTL    is defined as
   \[
\varphi 
::= 
true  
\mid a 
\mid \neg a
\mid \varphi_1\wedge \varphi_2
\mid \bigcirc \varphi
\mid \varphi_1 U \varphi_2,
\] 
where $a \in AP $ is an atomic proposition;
$\neg$ and $\wedge$ are Boolean operators ``negation'' and ``disjunction'', respectively; 
$\bigcirc$ and $U$ are temporal operators ``next'' and ``until'', respectively. 
One can also induce  temporal operators such as 
``eventually''  by $\lozenge \varphi =true U \varphi$.    
Note that, in scLTL, negation can only be applied in front of atomic propositions. 
Therefore, one cannot define temporal operator ``always''  $\square\varphi=\neg\lozenge\neg\varphi$ as the case of general LTL. 

In general, LTL formula are evaluated on infinite words over $2^{\mathcal{AP}}$. 
For any infinite word $\rho\in (2^{\mathcal{AP}})^\omega$, we denote by $\rho\models \varphi$ if word $\rho$ satisfies  LTL formula $\varphi$.
The reader is referred to \cite{baier2008principles} for details of the semantics of LTL. 
However, for an scLTL formula,  its satisfaction can be determined in finite horizon. It is well-known that, for any  infinite word $\rho=\rho_{0}\rho_{1} \dots \in (2^{\mathcal{AP}})^\omega$ 
such that  $\rho\models \varphi$, it has a  \emph{finite good prefix}  $\rho_{pref}=\rho_{0}\rho_{1} \dots\rho_n$ in the sense that $\rho_{pref}\rho'\models \varphi$ for any $\rho'\in  (2^{\mathcal{AP}})^\omega$.  
We denote by $\textsf{Word}_{pref}(\varphi)$ the set of all finite good prefixes for scLTL formula $\varphi$, 
and for a finite word $\rho\in (2^{\mathcal{AP}})^*$, we also write $\rho \models \varphi$ if $\rho\in \textsf{Word}_{pref}(\varphi)$. For system $G$, we denote by $G\models \varphi$ if 
every infinite path in $G$ has a finite good prefix. 
A good prefix is said to be minimal if it has no good prefix except itself, and we denote by $\textsf{Word}_{pref}^{min}(\varphi)$ the set of minimal good prefixes. 
Intuitively,  $\textsf{Word}_{pref}^{min}(\varphi)$ represents all words that \emph{exactly accomplishes the task} $\varphi$ for the first time. 
 We denote by $\sigma \models_1 \varphi$   if $\sigma \in \textsf{Word}_{pref}^{min}(\varphi)$, i.e., $\sigma$ exactly accomplishes the task for the first time. 
 Therefore,  $\sigma \not\models_1 \varphi$ means either the task is not yet satisfied or has been satisfied before.

The set of words satisfying an scLTL formula can be accepted by a \emph{deterministic finite-state automata}  (DFA). Formally,  a DFA  is  a 5-tuple  
\[
A = (S,s_0,\xi,\Sigma,F),
\]
where $S$ is the set of states, 
$s_0 \in S$ is the initial state,
$\Sigma$ is the alphabet,
$\xi: S \times \Sigma \rightarrow S$ is the deterministic transition function, 
and $F \subseteq S$ is the set of accepting states. 
The transition function  can be extended to $\xi: S\times \Sigma^{*}\to S$ recursively by: 
$\forall x \in S, \rho \in \Sigma^{*}, \sigma\in \Sigma, \xi(x,\rho\sigma)=\xi(\xi(x, \rho), \sigma)$. 
We denote by $\mathcal{L}(A)$ the set of all finite words \emph{accepted} by $A$, i.e., 
$\mathcal{L}(A)=\{\rho \in \Sigma^*:  \xi(s_0,\rho)\in F\}$. 
  
For any scLTL formula  $\varphi$, there always exists a DFA $A_\varphi$ over $\Sigma = 2^{\mathcal{AP}}$ that only accepts all  good prefixes, i.e.,  $\mathcal{L}(A_\varphi) = \textsf{Word}_{pref}(\varphi)$.
To distinguish the exact satisfaction of the task for the first time and after the first satisfaction, we  further modify DFA $A_\varphi$ by the following steps: (1) add  a new accepting state $s_F$ with self-loops for all labels in $\Sigma = 2^{\mathcal{AP}}$;  and   
(2) remove all output transitions from the original accepting states $F$, and add new transitions from $F$ to $s_F$ with labels for all $\Sigma = 2^{\mathcal{AP}}$;   and 
(3)  add a new ``bad'' state $s_B$, and for each state $s$, if any $\sigma \in 2^{\mathcal{AP}}$ is undefined, then we add a transition from $s$ to $s_B$ labeled by $\sigma$.  
Clearly, the modified DFA generates 
all words $(2^{\mathcal{AP}})^*$ and accepts exactly the same words as the original DFA.
Hereafter, we will use this modified version and still denote it by $A_\varphi$.

\subsection{Controller}
Since not all traces in  the system satisfy the scLTL formula, 
a feedback \emph{controller} (or policy) should be imposed. 
We consider a partially-observed setting, i.e., the controller cannot obtain the perfect state information of the system. To this end, we consider an observation mapping 
\[
H: X \rightarrow O,  
\]
where $O$ is the set of observation symbols and for each $x\in X$, $H(x)$ represents the observation of the controller at state $x$.  
Then for each path $\tau=x_{0}\dots x_{n}$, its corresponding observation is $H(\tau)=H(x_0)\dots H(x_{n})$. We define 
$H(\textsf{Path}(G))=\{ H(\tau) \in O^{*} :  \tau  \in \textsf{Path}(G)\}$ the set of all observations.
Then a controller for $G$ is a function 
\[
C : H(\textsf{Path}(G)) \rightarrow U,
\]
which determines the control input based on the observation.
The closed-loop system under controller $C$ is denoted by $G_{C}$. Specifically, we say that a path $ \tau = x_{0} \xrightarrow{u_1} x_{1} \xrightarrow{u_2} \cdots \xrightarrow{u_n} x_{n}$ is feasible under $C$ if $\tau \in \textsf{Path}(G)$ and $u_{i} = C(H(x_{0}\dots x_{i-1})),\forall i\in\{1,\dots,n-1\}$. We denote by $\textsf{Path}(G_{C})$ and $\textsf{trace}(G_{C})$ the set of all feasible paths and traces in $G$ under controller $C$, respectively.  
For technical purposes, we assume that for any two states having the same observation, 
they have the same set of active control inputs, 
i.e., $\forall x,x'\!\in\! X: H(x)\!=\!H(x')\Rightarrow U(x)\!=\!U(x')$.



\section{Intention-Aware Secure Synthesis Problem}

\subsection{Intruder Model}
During the execution of the system, the information available to the controller may also be released to the outsider due to, for example, insecure communications or malicious attacks. 
In this work, we consider an attacker modelled as a passive observer having the following capabilities:\vspace{-3pt}
\begin{itemize}
    \item 
    It knows the system model $G$, scLTL task $\varphi$, as well as the functionality of the controller $C$; and  
    \item 
    It can also access the observation of the controller. \vspace{-3pt}
\end{itemize}
Therefore, for each path $\tau=x_{0}\dots x_{n}$,  the observation of the intruder is also $H(\tau)=H(x_0)\dots H(x_{n})$.

\begin{remark}
The above intruder model has been widely considered in both computer science and control engineering literature; see, e.g., \cite{lafortune2018history,liu2022secure}. This model is motivated by the fact that offline designed policies are usually public information, while actual online trajectories are usually only partially released to the outsider. Therefore, the intruder needs to use this partial online information together with the offline model  information  to infer the behavior of the system. 
\end{remark}

\subsection{Unpredictable Security Requirement}

Depending on what information the system wants to hide, the security of the system can be defined differently. 
In this work, we consider the following security requirement:
\begin{itemize}
    \item 
    the system does not want the intruder to infer  too early 
    \emph{when it will accomplish the task}.
\end{itemize}
In other words, the controlled behavior of the system needs to be \emph{unpredictable} in the sense that, by observing external information-flow, the intruder is unable to infer confidentially that the robot will finish task exactly after a certain number of steps.

The above requirement was originally formulated by the notion of \emph{$K$-step pre-opacity} in~\cite{yang2022securepre}, where the completion of the task is modelled by reaching a target state. We then formulate our unpredictable controller synthesis problem by adopting this notion. First, we introduce the notion of $K$-step unpredictability w.r.t.\ an scLTL task.

\begin{mydef}[${K}$-Step Unpredictability]\label{def:unpre}
Given  system $G$, scLTL task $\varphi$ and controller $C$ with output function $H$, a finite path $\tau \in \textsf{Path}(G_{C})$ is said to be $K$-step unpredictable w.r.t.  $H$ and $\varphi$ if 
\begin{align}
    &(\forall m\ge K)(\exists \tau_{1}\tau_{2} \in \textsf{Path}(G_{C}))\text{ s.t. }\\
    &[|\tau_{2}|=m]\wedge [H(\tau)=H(\tau_{1})] \wedge [\textsf{trace}(\tau_{1}\tau_{2}) \not\models_1 \varphi].\nonumber
\end{align} 
We say that the controlled system $G_C$ is $K$-step unpredictable if all paths in it are $K$-step unpredictable.
\end{mydef}

Intuitively, $K$-step unpredictability   requires that for any feasible path $\tau$ in $G_C$ and any future instant $m\ge K$, there exists another observation-equivalent path $\tau_1$, such that after $\tau_1$, the system will not accomplish the task for the first time in  exact   $m$-steps. With this notion, we formulate the synthesis problem that we solve in this paper. 

 \begin{myprob}[Unpredictable Control Synthesis Problem] \label{problem1}
Given system $G$,  scLTL task $\varphi$ and output function $H$ for both intruder and controller, determine a controller $C$ such that 
(i) $G_C$ is live; and (ii) $G_C\models \varphi$; and 
(iii) $G_C$ is $K$-step unpredictable.  
 \end{myprob}

\begin{remark}
In the above problem formulation, 
we require that the controlled system $G_C$ is live, i.e., it needs to work indefinitely. In practice, since we consider scLTL task here, the robot can stop once it knows for sure that the task has been completed.  However, this assumption is mainly for the sake of simplification of analysis and is without loss of generality. For example, one can add a  virtual ``stop'' state with self-loops to mimic liveness when the robot stops. 
\end{remark}

\subsection{Product System and Problem Transformation}
To incorporate the task information into the system model, we construct the \emph{product} of system and the task DFA. 

\begin{mydef}[Product System]
Given system $G$ and scLTL task $\varphi$, 
let $A_\varphi = (S, s_0, \xi, \Sigma, F\cup \{s_F\})$ be the (modified) DFA such that  $\mathcal{L}(A_\varphi)=\textsf{Word}_{pref}(\varphi)$. The \emph{product} of $G$ and $A_\varphi$ is a new transition system 
\[
\tilde{G}= ( \tilde{X}, \tilde{x}_0, U, \rightarrow_\otimes, \mathcal{AP}, \tilde{L}),
\]
where  
$\tilde{X} \subseteq X \times X_{F}$ is the set of states;
$\tilde{x}_0=(x_{0},\xi(s_0,L(x_{0}))) \in \tilde{X}$ is the initial state;
$U$ is the set of control inputs;
$\rightarrow_\otimes \subseteq \tilde{X} \times U \times \tilde{X}$ is transition relation defined by:   
for any $\tilde{x}_1=(x_1,s_1)\in \tilde{X}, \tilde{x}_2=(x_2,s_2)\in \tilde{X}$ and  $u\in U$,
we have $\tilde{x}_1   \xrightarrow{u}_\otimes \tilde{x}_2$ if 
(i) $x_{1}  \xrightarrow{u} x_{2}$ and 
(ii) $s_{2}\!=\!\xi(s_{1},L(x_{2}))$; 
$\mathcal{AP}$ is the same set of atomic propositions; and 
$\tilde{L}: \tilde{X} \rightarrow 2^{\mathcal{AP}}$ is labeling function such that, 
for any $ \tilde{x}=(x,s)$, we have $\tilde{L}(\tilde{x})=L(x)$.
\end{mydef}

Intuitively, $\tilde{G}$ is obtained by incorporating the information of $A_\varphi$ into system model $G$. Since the modified $A_\varphi$ generates all words, we have $\textsf{Trace}(\tilde{G})=\textsf{Trace}(G)$. 
Furthermore, since $A_\varphi$ is deterministic, the path in $\tilde{G}$ is completely determined by $G$, i.e., $\textsf{Path}(\tilde{G})=\textsf{Path}(G)$.  
Also, we define observation mapping  $\tilde{H}:\tilde{X} \rightarrow O$ by: for $(x,s)=\tilde{x} \in \tilde{X}$, we have $\tilde{H}(\tilde{x})=H(x)$.
Then any controller $C$ designed for $\tilde{G}$ can be applied to $G$, and vice versa. Hereafter, we will focus on the product system $\tilde{G}$.


Note that, the information of the satisfaction of scLTL task $\varphi$ has also been encoded in the second component of $\tilde{G}$. To see this more clearly, we define
\begin{align}
\tilde{X}_{F} =& \{ (q,s) \in \tilde{X} : s \in F  \},\nonumber\\
\tilde{X}_{F \cup \{s_F \}} =& \{ (q,s) \in \tilde{X} : s \in F \cup \{s_F \} \}.\nonumber
\end{align} 
Then for any path $\tau \in \textsf{Path}(\tilde{G})$, we have the followings: 
\begin{align}
   \textsf{trace}(\tau) \models_1 \varphi &\Leftrightarrow  \textsf{last}(\tau) \in \tilde{X}_{F},\\
      \textsf{trace}(\tau) \models \varphi &\Leftrightarrow  \textsf{last}(\tau) \in \tilde{X}_{F \cup \{s_F \}}. \label{eq:task}
\end{align}
Intuitively, states set $\tilde{X}_F$ captures the exact satisfaction of scLTL task $\varphi$ in $\tilde{G}$. 
We also call $\tilde{X}_F$  \emph{secret states} since they represent those states whose visits do not want to be predicted by the intruder. 
Therefore, the unpredictability requirement w.r.t.\ the scLTL task $\varphi$ can be equivalently expressed in terms of the unpredictability of reaching secret states. 
In addition to replace $\textsf{trace}(\tau_{1}\tau_{2}) \not\models_1 \varphi$ in Definition~\ref{def:unpre} by $\textsf{last}(\tau_{2}) \notin  \tilde{X}_{F}$, 
here, we provide a stronger result showing that the requirement of ``$\forall m\geq K$'' can also be simplified by only considering $m=K$.





\begin{mypro} \label{prop:trans}
Given  system $\tilde{G}$ and  controller $C$, $\tilde{G}_{C}$ is $K$-step unpredictable w.r.t. $H$ and $\varphi$  if and only if for any finite path $\tau \in \textsf{Path}(\tilde{G}_{C})$
there exists a path $\tau_{1}\tau_{2} \in\textsf{Path}(\tilde{G}_{C})$ s.t.\ 
\[
[|\tau_{2}|=K]\wedge [\tilde{H}(\tau)=\tilde{H}(\tau_{1})] \wedge [\textsf{last}(\tau_{2}) \notin \tilde{X}_{F}].
\]
\end{mypro}
\begin{pf}
Due to space constraint, all proofs in this paper are omitted. The complete proofs are available in {\url{https://xiangyin.sjtu.edu.cn/ifac-pf.pdf}}.
\end{pf}

Based on the product system $\tilde{G}$ and  Proposition~\ref{prop:trans}, 
we can transform Problem~1 regarding system $G$ and scLTL task $\varphi$ to a problem depending only on $\tilde{G}$. 
That is, our objective beomces to synthesize a live controller $C$ for $\tilde{G}$ such that  all traces in $\tilde{G}_C$ end up with $\tilde{X}_{F}$   are $K$-steps unpredictable as characterized by  Proposition~\ref{prop:trans}. 
 To simplify the notation, in the rest of this paper, we  denote   system $\tilde{G}= ( \tilde{X}, \tilde{x}_0, U, \rightarrow_\otimes , \mathcal{AP}, \tilde{L})$ by $G = ( X, x_{0}, U, \rightarrow , \mathcal{AP}, L)$ by understanding how the product is constructed. 
 We also denote states set $\tilde{X}_F$ and $\tilde{X}_{F \cup \{s_F \}}$ by $X_F$ and $X_{F \cup \{s_F \}}$, respectively.

\section{Prediction Sets and Bipartite Transition Systems}

\subsection{Prediction Sets} 
Compared with the existing security-aware controller synthesis problems for current-state opacity, our $K$-step unpredictable controller synthesis problem has the following new challenge. 
In the synthesis for current-state opacity, whether or not the secret of the system is revealed to the intruder can be completely determined by the historical information up to the current instant. However, in our problem, whether or not the system is secure depends on its behavior \emph{in the future}. This is not a verification problem for open-loop system since all future behaviors  are given and fixed. However, in the synthesis problem, the future behaviors of the system are unknown and  depend on the control decisions in the future, which have not yet been determined. This \emph{future dependency issue} makes our synthesis problem particularly challenging. 

To resolve the future dependency issue, we propose a novel approach by ``borrowing'' information in the future. The general idea is as follows. At each instant, the system makes a \emph{prediction} regarding in how many number of steps, the scLTL task will be accomplished, or equivalently, secret states $X_F$ will be reached. In order to make the prediction valid, the system should accomplish the task as it predicted; otherwise, such a prediction will be considered invalid and be truncated.

To formalize the above idea, we define a \emph{prediction} as a $(K+1)$-dimensional binary vector $h=(h{[0]},h[1],\cdots,h[K])$ $ \in \{0,1\}^{K+1}$, where for each $i=0,1,\dots,K$, \vspace{-3pt}
\begin{itemize}
    \item 
    $h[i]=0$  means that the system \emph{may not reach} secret states $X_F$ in exactly $i$ steps; and 
    \item 
    $h[i]=1$  means that the system \emph{will reach} secret states $X_F$ \emph{for sure} in exactly $i$ steps.\vspace{-3pt}
\end{itemize}
We denote by $\mathcal{H}=\{0,1\}^{K+1}$ the prediction set. We augment the state space of $G$ with prediction set and denote by 
$\hat{X} = X\times \mathcal{H}$ 
the augmented state space.  
For each augmented state $\hat{x}=(x,h)\in \hat{X}$, 
we define $\textsf{state}(\hat{x})=x$ and $\textsf{pred}(\hat{x})=h$ as its state and prediction components, respectively. 
Then, for a set of augmented states 
$\imath= \{(x_{1},h_{1}),...,$
 $(x_{n},h_{n}) \} \subseteq\hat{X}$, we define 
$\textsf{state}(\imath)=\{x_{1},...,x_{n}\}$ and $\textsf{pred}(\imath)=\{h_{1},...,h_{n}\}$.  With some abuse of notation, for $(x,h)=\hat{x} \in \hat{X}$, we also write $H(\hat{x})=H(x)$.

Now, we discuss what properties a ``good'' prediction should have. First, in an augmented state $\hat{x}=(x,h)\in \hat{X}$, its prediction for the current instant, i.e., $h[0]$, should be consistent with the fact of the current state $x$.  
This is captured by the following definition. 

\begin{mydef}[Current Consistency]
An augmented state $\hat{x} \in \hat{X}$ is called \emph{currently consistent} if
\begin{equation}
   \textsf{pred}(\hat{x})[0]=1\Leftrightarrow 
   \textsf{state}(\hat{x})\in X_F.
\end{equation}
We denote by $\hat{X}_{cons}\subseteq \hat{X}$ the set of all currently consistent augmented states.
\end{mydef} 

Second, let $\hat{x}\in \hat{X}$ be an augmented state, and $\imath\in 2^{\hat{X}}$ be a set of augmented states representing all successor states of $\hat{x}$ in \emph{one step}. Then the predictions of $\imath$ at the next instant should be consistent with the predictions of $\hat{x}$ at the current instant in the sense that \vspace{-3pt}
\begin{itemize}
    \item 
    if $\hat{x}$ predicts that the system will reach secret states for sure at some instants, then \emph{all of} its successor states $\imath$ should agree on this prediction;  
    \item 
    if $\hat{x}$ predicts that the system may not reach secret states at some instants, then \emph{some of} its successor states in $\imath$ should maintain this possibility. \vspace{-3pt}
\end{itemize}
Formally, we have the following definition.

\begin{mydef}[One-Step Consistency]
Let $h\in \mathcal{H}$ be a prediction  and $H\in 2^{ \mathcal{H}}$ be a set of predictions. We say $h$ and $H$ are  \emph{one-step consistent} if for each $i=1,\dots,K$,  
\begin{align}
   h[i]\!=\!1&\ \Rightarrow \  \  \forall h'\in H:h'[i-1]\!=\!1,  \nonumber\\
   h[i]\!=\!0&\ \Rightarrow \ \  \exists  h'\in H:h'[i-1]\!=\!0. 
\end{align}
We denote by $(h,H)\in \Delta$ if they are one-step consistent.
\end{mydef}

Due to the partial observability, the system may not know the precise state and the prediction. Instead, it may hold a \emph{belief} regarding the augmented state.  

\begin{mydef}[Belief States]
A \emph{belief state} is a  set of currently consistent augmented states  $\imath\in 2^{\hat{X}_{cons}}$ such that
\[
\forall (x,h), (x',h')\in \imath: x\!=\!x'\Rightarrow h\!=\!h'.
\]
We denote by $\mathcal{B}\subseteq 2^{\hat{X}_{cons}}$ the set of all belief states. 
\end{mydef} 
Intuitively, in a belief state, the predictions for the same state are the same, i.e., we have $|\textsf{state}(\imath)|=|\imath|$. 
If all   states in a belief state has the same prediction, then even if the belief state is not a singleton, we can still make a conclusion regarding when the secret state will be reached. 
\begin{mydef}[Secure Belief States]
A  belief state $\imath\in \mathcal{B}$ is said to be \emph{insecure} if 
\[
\forall \hat{x} \in \imath:  \textsf{pred}(\hat{x})[K]=1.
\]
Otherwise, $\imath$ is called \emph{secure}. 
We denote by $\mathcal{B}_{ins}$ and $\mathcal{B}_{sec}$ the set of insecure and secure belief states, respectively.  
\end{mydef}

\subsection{Bipartite Transition Systems}
In order to synthesize an unpredictable controller, our approach is to enumerate all possible control actions over the belief state space. Then, we solve a safety game to avoid those insecure belief states at which the intention of accomplishing the task in $K$ steps is revealed.  

 \begin{mydef}(Bipartite Transition Systems)\label{def:BTS}
A bipartite transition system (BTS) of $G$ is a 7-tuple
\[
T=(Q_Y,Q_Z,\delta_{YZ},\delta_{ZY},U,O,Y_0),
\]
where\vspace{-3pt}
\begin{itemize}
    \item 
    $Q_Y \subseteq  \mathcal{B}_{sec} \times O$ is the set of $Y$-states; 
    \item
    $Q_Z\subseteq \mathcal{B}_{sec}\times U$ is the set of $Z$-states; 
    \item 
    $\delta_{YZ}: Q_Y \times U \rightarrow 2^{Q_Z}$ is a non-deterministic transition function from $Y$-states to $Z$-states defined by: for any $y=(\imath,o)\in Q_Y, u\in U$, and $z=(\imath',u')\in Q_Z $, we have 
    $z\in \delta_{YZ}(y,u)$ if
    (i)  $u=u'$; and
    (ii) $\textsf{state}(\imath')
        =\mathbf{NX}_u(\textsf{state}(\imath))$; and
    (iii)
        for $\hat{x}\in \imath$, 
        we have $(\textsf{pred}(\hat{x}),\textsf{pred}(\imath'(\hat{x},u)))\in \Delta$, where  
        \[
        \imath'(\hat{x},u)  =\{ \hat{x}' \in \imath' : \textsf{state}(\hat{x}') \in \mathbf{NX}_{u}(\textsf{state}(\hat{x}))  \}.
        \]     
    \item 
    $\delta_{ZY}: Q_Z \times O \rightarrow Q_Y$ is the deterministic transition function from $Z$-states to $Y$-states defined by: 
    for any $z=(\imath,u) \in Q_{Z},o  \in O$, and $y=(\imath',o')\in Q_Y$, we have $\delta_{ZY}(z,o)\!=\!y$
    if 
    (i) $o'\!=\!o$; and (ii) 
        $\imath'\!=\!\{ \hat{x} \!\in\! \imath\!:\! H(\hat{x})\!=\!o \}$. 
    \item 
    $U$ is the set of input;
    \item 
    $O$ is the output set;
    \item 
    $Y_0= \{(\{ (x_0,h)  \},H(x_0)): (x_0,h) \in \mathcal{B}_{sec}\}$ is the  set of possible initial $Y$-states.
\end{itemize}
\end{mydef}

Intuitively, 
for a $Y$-state $(\imath,o)$,  
$\textsf{state}(\imath)$ represents states estimation of controller/intruder, 
$\textsf{pred}(\imath)$ is its corresponding predictions, and 
$o$ is the last observation. 
A transition from a $Y$-state to a $Z$-state represents that controller chooses a control input and every prediction of system state in $Y$-state is one-step consistent with prediction of its next step system states set in $Z$-state. A transition from $Z$-state to $Y$-state simply restricts the belief based on the observation.  
Note that for $(\imath,o)\in Q_{Y}$ and $(\imath',u)\in Q_{Z}$, $o$ and $u$ are redundant information since they are uniquely determined by their input transitions, and we use them mainly to distinguish between $Y$ and $Z$-states. With some abuse of notation, for $q=(\imath,b) \in Q_{Y} \cup Q_{Z}$, 
we sometimes omit $b$ and  also write $\textsf{state}(q)=\textsf{state}(\imath)$ and $\textsf{pred}(q)=\textsf{pred}(\imath)$.

For the purpose of control, we need to further require that, in each $Y$-state, there exists a control input under which it has successor and, in each $Z$-state all feasible observations are defined. Formally, 
given a BTS $T$, we say \vspace{-3pt}
\begin{itemize}
    \item 
    a $Y$-state $y=(\imath,o)\in Q_Y$ is \emph{complete} if there exists a control input  defined at $y$ in $T$, i.e.,  
    $\exists u\in U: \delta_{YZ}(y, c) \neq \emptyset$; 
    \item 
    a $Z$-state $z=(\imath,u)\in Q_Z$ is \emph{complete} if  all feasible observations are defined at $z$ in $T$, i.e.,  
    $\forall \hat{x}\in \imath: \delta_{ZY}(z,H(\hat{x}))!$.
\end{itemize} 
Then BTS $T$ is said to be complete if all $Y$-states and $Z$-states in it are complete. 

Note that a complete BTS $T$ may contain multiple controlled behaviors since 
(i) multiple control inputs may be defined at the same $Y$-state; and 
(ii) even if a specific control input is selected at a $Y$-state, it may lead to different  $Z$-state corresponding to different predictions.
In order to ``decode'' a controller from $T$, we need to resolve the above nondeterminism.

\begin{mydef}[Deterministic BTS]
A complete BTS $T=(Q_Y,Q_Z,\delta_{YZ},\delta_{ZY},U,O,Y_0)$ is said to be  \emph{deterministic} if \vspace{-3pt}
\begin{itemize}
    \item[(i)] 
    The initial $Y$-state is unique, i.e., 
    $Y_0=\{y_0\}$; and  
    \item[(ii)]
    For any $Y$-state, there exists a unique control input  defined, i.e., 
    $\forall y\in Q_Y: |\{u:\delta_{YZ}(y,u)!\}|=1$; and 
    \item[(iii)]
    For each $Y$-state and control input, the transition is deterministic, i.e., 
    $\forall y\in Q_Y,u\in U: |\delta_{YZ}(y,u)|\leq 1$.
\end{itemize}
\end{mydef}

For a deterministic BTS $T$,  we denote by  $U_T(y)$ the unique control input defined at  $Y$-state $y\in Q_Y$ in $T$. 
Let $T=(Q_Y,Q_Z,\delta_{YZ},\delta_{ZY},U,O,y_0)$ be a deterministic BTS and $\pi=o_0u_0o_1u_1\dots o_{n-1}u_{n-1}o_n\in O(UO)^*$ be an alternating sequence of observations and control inputs. 
Then $\pi$ visits a unique sequence of states 
$y_0 z_0 y_1 z_1\dots y_{n-1} z_{n-1} y_{n}$, where $y_0$ is the initial $Y$-state, and for each $i\in\{0, 1, \dots, n-1\}$, we have $z_i=h_{ZY}(y_i,u_i)$ and
$y_{i+1}=h_{ZY}(z_{i},o_{i+1})$. 
We denote $Y_T(\pi)=y_n$ as the last $Y$-state reached by 
$\pi$. 
Note that, since $T$ is deterministic, the above state sequence is uniquely determined by its observation part $\mathbf{o}=o_0o_1\dots o_n$
Therefore, we can also write  $Y_T(\mathbf{o})$ as the unique $Y$-state reached by a sequence whose observation part is $\mathbf{o}$.  
Then for a deterministic BTS $T$,  we can ``decode'' a controller, denoted by $C_T$, as follows: 
for any observation $\mathbf{o}\in H(\textsf{Path}(G))$, we have
\begin{equation}
    C_T(\mathbf{o})=U_T(Y_T(\mathbf{o})),
\end{equation}
which is the unique control input defined at the unique $Y$-state reached by $\mathbf{o}$.

\subsection{Properties of the BTS}
Given a controller $C$, for any observation sequence 
$\mathbf{o}\in H(\textsf{Path}(G_C))$, we define the \emph{state estimate} (for both controller and the intruder) upon $\mathbf{o}$ by  
\begin{equation}
    \mathcal{E}_{C}(\mathbf{o})=\{\textsf{last}(\tau) \in X  : \tau \in \textsf{Path}(G _{C})\wedge H(\tau)= \mathbf{o}\}.
\end{equation}
The state estimate $\mathcal{E}_{C}(\mathbf{o})$ essentially includes all possible current states of the controlled system by  observating $\mathbf{o}$. The following result states that, for a BTS induced controller, the \emph{state estimation} upon $\mathbf{o}$ is essentially the $Y$-state reached by observation sequence $\mathbf{o}$ in the BTS.
\begin{mypro} \label{5-16}
Let $T$ be a deterministic BTS and $C_T$ be its induced controller. 
Then, for any $\tau \in \textsf{Path}(G_{C_T})$,  
\begin{equation}
\textsf{state}( Y_T(H(\tau) )=\mathcal{E}_{C_T}( H(\tau) ).
\end{equation}
\end{mypro}

Since BTS is complete, for $\tau \in \textsf{Path}(G_{C_{T}})$, $Y_T(H(\tau)$ has successor $Z$-state. Therefore, $\tau$ also has successor states. The above result further indicates that, given controller $C_T$ induced by  deterministic BTS, $G_{C_T}$ is live.

Next, we show that for a deterministic BTS, the prediction of each state indeed corresponds to the information of when it will reach a secret state $X_F$, i.e., complete the task, under the induced controller. To this end, 
for any  path $\tau\in \textsf{Path}(G_{C})$, we define
\[
 \textsf{Reach}_{G_{C}}^i(\tau)
 =\{ \textsf{last}(\tau\tau') \in X : \tau \tau' \in \textsf{Path}(G_{C}), |\tau'|=i \}  
\]
as the set of states the controlled system $G_{C}$ can reach in $i$ steps following $\tau$.
Then, we have the following result. 
\begin{mypro}\label{5-9}
Let $T$ be a  deterministic BTS and $C_T$ be its induced controller. 
For path $\tau\in \textsf{Path}(G _{C})$, let 
$(\imath,o)=Y_T(H(\tau))$ be the $Y$-state reached in $T$ along the observation of $\tau$. Then we have  \vspace{-3pt}
\begin{itemize}
    \item[(i)]
    there exists a unique augmented state $\hat{x}\in \imath$ such that $\textsf{state}(\hat{x})=\textsf{last}(\tau)$; and 
    \item[(ii)]
    for the unique  augmented state $\hat{x}\in \imath$,    we have
    \[
    \forall i\!=\!\{0,\dots,K\}:   \textsf{pred}(\hat{x})[i]\!=\!1
    \Leftrightarrow
    \textsf{Reach}_{G_{C_T}}^i(\tau)\!\subseteq\! X_F.
    \]
\end{itemize}  
\end{mypro}

The above result shows that the prediction part of each augmented state indeed captures the future behavior of the system. Furthermore, in the definition of BTS, we only consider secure belief states. This means that, under any BTS-induced controller, the intruder can never predicts that the system will reach $X_F$ more than $K$ steps ahead. This leads to the following result.  

\begin{mypro} \label{5-10}
Let $T$ be a   deterministic BTS and $C_T$ be its induced controller. 
Then the controlled system $G_{C_T}$ is $K$-step unpredictable. 
\end{mypro}

\section{Controller Synthesis Procedure}

In the previous section, we have shown how to ``decode''  an unpredictable controller from a deterministic BTS. In this section, we discuss how to build such a deterministic BTS. 
Our approach consists of two steps: \vspace{-3pt}
\begin{itemize}
    \item 
    First, we build the largest complete BTS, called the all enforcement structure (AES), as the solution space for all unpredicable controllers; 
    \item 
    Second, we extract a deterministic BTS from the AES by solving a reachability game such that the scLTL task is also enforced.
\end{itemize}

First, we introduce the notion of all enforcement structure.

\begin{mydef}[All Enforcement Structure]
Given system $G$, its \emph{all enforcement structure} 
$AES(G)=(Q^{AES}_Y,Q^{AES}_Z,$ $\delta^{AES}_{YZ}, \delta^{AES}_{ZY},U,O,Y^{AES}_0)$ is defined as the largest complete BTS. 
By ``largest'', we mean that for any complete BTS $T$, we have that $T \sqsubseteq AES(G)$, where $\sqsubseteq$ denotes the standard sub-system (graph) inclusion. 
\end{mydef}

The AES can be constructed as follows. 
First, starting from each possible initial $Y$-state, 
we enumerate (i) all possible control inputs at each $Y$-state, 
(ii) all possible predictions for each $Y$ to $Z$ transitions, and 
(iii) all possible observations. 
That is, all transitions satisfying the requirements in Definition~\ref{def:BTS} are defined.
Note that, the resulting BTS may be incomplete since some transitions may lead to inconsistent beliefs or insecure beliefs. 
Therefore, we need to further remove incomplete $Y$ or $Z$-states iteratively until the BTS becomes complete. Then, the resulting structure is the AES. 

Next, we need to extract a deterministic BTS $T$ from the AES such that  scLTL task is enforced. Since our purpose is to reach states in $X_{F \cup \{ s_{F} \}}$, 
we define 
\begin{align}
 Q_{Y}^{F(0)}=& \{ y \in Q_{Y}^{AES} \mid \textsf{state}(y) \subseteq X_{F \cup \{ s_{F} \}} \},\nonumber\\
 Q_{Z}^{F(0)}=& \{ z \in Q_{Z}^{AES} \mid   \textsf{state}(z) \subseteq X_{F \cup \{ s_{F} \}} \},\nonumber\
\end{align}
as the sets of $Y$ and $Z$ states such that their state estimate components are subsets of $X_{F \cup \{ s_{F} \}}$, respectively.  
Therefore,  our goal is to choose an initial $Y$-state from the AES, and choose a control input and a prediction for each $Y$-state, such that $Q_{Y}^{F(0)}\cup Q_{Z}^{F(0)}$ is reached in a finite number of steps. This is essentially a reachability game on the AES by considering all $Y$-states as control nodes and all $Z$-states as adversary nodes. To this end, 
we define $Q_Y^{F(k)}$ and $Q_Z^{F(k)}$ as follows: 
\begin{align}
    Q_Y^{F(k+1)}=& \{y\!\in\! Q^{AES}_Y: \exists u\!\in\! U\text{ s.t. }\delta_{YZ}^{AES}(y,u)\!\cap\! Q_Z^{F(k)}\!\neq\! \emptyset \}\nonumber\\
     &\cup Q_Y^{F(k)},\\
   Q_Z^{F(k+1)}=& \{z\!\in\! Q^{AES}_Z: \forall o\!\in\! O\text{ s.t. }\delta_{ZY}^{AES}(z,o)\!\subseteq\! Q_Y^{F(k)} \}\nonumber\\
     &\cup Q_Z^{F(k)}.
\end{align}
Define $Q_Y^{F}=\bigcup_{k\geq 0} Q_Y^{F(k)}$ and $Q_Z^{F}=\bigcup_{k\geq 0} Q_Z^{F(k)}$.  
Intuitively, for each $Y$-state $y$, 
$y\in Q_Y^{F(k)}\setminus Q_Y^{F(k-1)}$ means that the controller can ensure to reach $X_{F \cup \{ s_{F} \}}$ in $k$ steps. 
Then we define a distance function 
$\textsf{dist}: Q_{Y}^{AES}\cup Q_{Z}^{AES} \to \mathbb{N}\cup \{\infty\}$ by:  
for each $Y$-state $y\in Q_Y^{AES}$,
\begin{equation} \label{equ:pick}
\textsf{dist}(y)\!=\!
 \left\{ 
     \begin{array}{l l} 
       k &, \text{if } y\in Q_Y^{F(k)}\setminus Q_Y^{F(k-1)} \\ 
       \infty &, \text{if } y\notin Q_Y^{F} 
     \end{array},
    \right. 
\end{equation}
and the same for each  $Z$-state $z\in Q_Z^{AES}$. 

\IncMargin{1em}
\begin{algorithm} 
\caption{Unpredictable Controller Synthesis}\label{alg:security} 
\KwIn{system $G$, scLTL formula $\varphi$}
\KwOut{controller $C_T$\vspace{3pt}}

construct DFA $A_\varphi$ for $\varphi$\\
construct product system of $G$ and $A_\varphi$\\
construct the AES $AES(G)$\\
compute distance function $\textsf{dist}$ by Equation~(\ref{equ:pick})\\
\eIf{$Y_0^{AES}\cap Q_Y^{F}=\emptyset$}
{\textbf{return} no solution exists}
{$T\gets \texttt{Extract}(AES(G), \textsf{dist} )$\\
\textbf{return} decoded controller $C_T$ from $T$}
\vspace{3pt}

\textbf{procedure} $\texttt{Extract}(AES(G),\textsf{dist} )$\\
pick  an initial  $y_0\in Y_0^{AES}\cap Q_Y^{F}$ and add $y_0$ into $T$\\
\While{$T$ is not complete}
{
\For{incomplete $Y$-state $y$ in $T$} 
{
find $y\xrightarrow{u}{z}$ in the AES s.t.\  $\textsf{dist}(z)<\textsf{dist}(y)$\\
add state $z$ to $Q_Z^T$ and transition $y\xrightarrow{u}{z}$ to $\delta_{YZ}^T$ 
}
\For{incomplete $Z$-state $z$ in $T$} 
{
add all successor $Y$-states and the associated transitions in the AES to $T$ 
} 
}
\end{algorithm}	

The complete synthesis algorithm is shown in Algorithm~\ref{alg:security}. 
After constructing the AES and the distance function $\textsf{dist}$, we extract a deterministic BTS from the AES by procedure \texttt{Extract}($\cdot$). Specifically, we start from an initial $Y$-state in $y_0\in Y_0^{AES}\cap Q_Y^{F}$; if none, then it means that the controller cannot ensure the completion of scLTL task $\varphi$. Next, at each $Y$-state, we choose a control input and a prediction to move to a $Z$-state such that the distance to target states are smaller. This procedure ensures that the scLTL task is accomplished in finite number of steps. Furthermore, by the construction of the AES, the induced controller is also unpredictable.   
The complexity of the algorithm is exponential in the size of the system. However, it is known that such an exponential complexity is unavoidable for partially-observed synthesis problem. 
The following theorem shows the correctness of the proposed algorithm.

\begin{mythm}  \label{thm:souandcom}
 Algorithm~\ref{alg:security} is  both sound and complete, i.e.,   its output is a solution to  Problem~\ref{problem1}, 
 and if it returns ``no solution'', then Problem~\ref{problem1} has no solution.
\end{mythm}

\section{Illustrative Case Study}
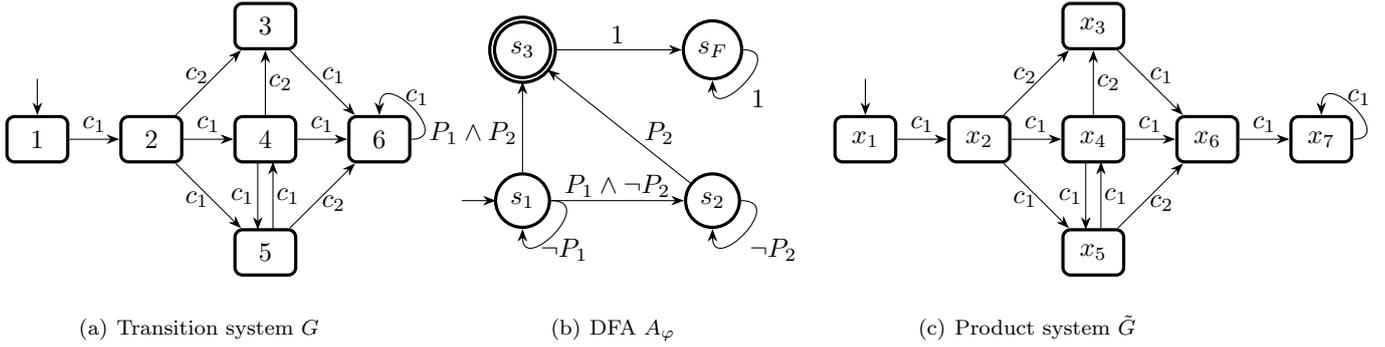
\begin{figure*}  
	\subfigure[Transition system $G$] 
	{\label{fig:LTS1} 
            \begin{minipage}[b]{0.28\linewidth}
               	\centering
    \begin{tikzpicture}
[
square/.style={rectangle, draw=black!255, fill=white!255, very thick, minimum height=6mm,minimum width=8mm,rounded corners = 1mm},
]
    \node [square](q1)at(0,0){$1$};
	\node [square](q2)at(1.5,0){$2$};
	\node [square](q3)at(3,1.5){$3$};
	\node [square](q4)at(3,0){$4$};
	\node [square](q5)at(3,-1.5){$5$};
	\node [square](q6)at(4.5,0){$6$};
	\node (c12)at(0.75,0.2){$c_{1}$};
	\node (c23)at(2.1,0.85){$c_{2}$};
	\node (c25)at(2.1,-0.8){$c_{1}$};
	\node (c24)at(2.25,0.2){$c_{1}$};
	\node (c36)at(3.9,0.85){$c_{1}$};
	\node (c43)at(3.2,0.75){$c_{2}$};
	\node (c45)at(2.7,-0.75){$c_{1}$};
	\node (c46)at(3.75,0.2){$c_{1}$};
	\node (c54)at(3.3,-0.75){$c_{1}$};
	\node (c56)at(3.9,-0.85){$c_{2}$};
	\node (c66)at(5,0.55){$c_{1}$};
\draw [-{Stealth}](0,0.8) --(0,0.32);
\draw [-{Stealth}](2.9,-0.3) --(2.9,-1.2);
\draw [-{Stealth}](3.1,-1.2) --(3.1,-0.3);
\draw[-{Stealth}] (q1) -- (q2);
\draw[-{Stealth}] (q2) -- (q4);
\draw[-{Stealth}] (q2) -- (q3);
\draw[-{Stealth}] (q2) -- (q5);
\draw[-{Stealth}] (q3) -- (q6);
\draw[-{Stealth}] (q4) -- (q3);
\draw[-{Stealth}] (q4) -- (q6);
\draw[-{Stealth}] (q5) -- (q6);
\draw[-{Stealth}] (q6) .. controls +(right:10mm) and +(up:10mm) .. (q6);
\end{tikzpicture} 
           \end{minipage}
	}
	\subfigure[DFA $A_\varphi$] 
	{\label{fig:moauto2} 
 \begin{minipage}[b]{0.28\linewidth}
	\centering
 \begin{tikzpicture}
[
acc/.style={circle, draw=black!255, fill=white!255, very thick, double,minimum height=6mm,minimum width=8mm,rounded corners = 1mm},
nor/.style={circle, draw=black!255, fill=white!255, very thick, minimum height=6mm,minimum width=6mm},
]
	\node [nor](q1)at(0,0){$s_{1}$};
	\node [nor](q2)at(2.5,0){$s_{2}$};
	\node [acc](q3)at(0,2){$s_{3}$};
	\node [nor](q4)at(2.5,2){$s_{F}$};

\draw[-{Stealth}] (-0.8,0) -- (-0.35,0);
\draw[-{Stealth}] (q1) -- (q3);
\draw[-{Stealth}] (q1) -- (q2);
\draw[-{Stealth}] (q2) -- (q3);
\draw[-{Stealth}] (q3) -- (q4);
\draw[-{Stealth}] (q1) .. controls +(right:10mm) and +(down:10mm) .. (q1);
\draw[-{Stealth}] (q2) .. controls +(right:10mm) and +(down:10mm) .. (q2);
\draw[-{Stealth}] (q4) .. controls +(right:10mm) and +(down:10mm) .. (q4);

\node (c12)at(1.25,0.2){$P_{1} \wedge \neg P_{2}$};
\node (c23)at(1.8,0.9){$ P_{2}$};
\node (c22)at(3.3,-0.65){$\neg P_{2}$};
\node (c11)at(0.55,-0.65){$\neg P_{1}$};
\node (c34)at(1.25,2.2){$1$};
\node (c44)at(3.1,1.35){$1$};
\node (c13)at(-0.65,0.9){$P_{1} \wedge P_{2}$};
\end{tikzpicture}
    \end{minipage}
	}	
	\subfigure[Product system $\tilde{G}$] 
	{ \label{fig:exm1}
 \begin{minipage}[b]{0.28\linewidth}
	\centering
     \begin{tikzpicture}
[
square/.style={rectangle, draw=black!255, fill=white!255, very thick, minimum height=6mm,minimum width=8mm,rounded corners = 1mm},
]
    \node [square](q1)at(0,0){$x_{1}$};
	\node [square](q2)at(1.5,0){$x_{2}$};
	\node [square](q3)at(3,1.5){$x_{3}$};
	\node [square](q4)at(3,0){$x_{4}$};
	\node [square](q5)at(3,-1.5){$x_{5}$};
	\node [square](q6)at(4.5,0){$x_{6}$};
	\node [square](q7)at(6,0){$x_{7}$};
	\node (c12)at(0.75,0.2){$c_{1}$};
	\node (c23)at(2.1,0.85){$c_{2}$};
	\node (c25)at(2.1,-0.8){$c_{1}$};
	\node (c24)at(2.25,0.2){$c_{1}$};
	\node (c36)at(3.9,0.85){$c_{1}$};
	\node (c43)at(3.2,0.75){$c_{2}$};
	\node (c45)at(2.7,-0.75){$c_{1}$};
	\node (c46)at(3.75,0.2){$c_{1}$};
	\node (c54)at(3.3,-0.75){$c_{1}$};
	\node (c56)at(3.9,-0.85){$c_{2}$};
	\node (c77)at(6.5,0.55){$c_{1}$};
	\node (c67)at(5.25,0.2){$c_{1}$};
	
\draw [-{Stealth}](0,0.8) --(0,0.32);
\draw [-{Stealth}](2.9,-0.3) --(2.9,-1.2);
\draw [-{Stealth}](3.1,-1.2) --(3.1,-0.3);
\draw[-{Stealth}] (q1) -- (q2);
\draw[-{Stealth}] (q2) -- (q4);
\draw[-{Stealth}] (q2) -- (q3);
\draw[-{Stealth}] (q2) -- (q5);
\draw[-{Stealth}] (q3) -- (q6);
\draw[-{Stealth}] (q4) -- (q3);
\draw[-{Stealth}] (q4) -- (q6);
\draw[-{Stealth}] (q5) -- (q6);
\draw[-{Stealth}] (q6) -- (q7);
\draw[-{Stealth}] (q7) .. controls +(right:10mm) and +(up:10mm) .. (q7);

\end{tikzpicture}   
    \end{minipage}
	} \vspace{-5pt}
\caption{For $\tilde{G}$, we have: $x_{1}=(1,s_{1})$, $x_{2}=(2,s_{2})$, $x_{3}=(3,s_{2})$, $x_{4}=(4,s_{2})$, $x_{5}=(5,s_{2})$, $x_{6}=(1,s_{3})$, $x_{7}=(7,s_{F})$.}
	\end{figure*}

In this section,  we revisit the motivating example in Section~\ref{sec:motivating-example} to illustrate our proposed approach. 

The mobility of the robot can be modeled as the non-deterministic  transition system shown in Fig.~\ref{fig:LTS1}. For instance, both $(2,c_{1},4)$ and $(2,c_{1},5)$ are legal transitions, which means that the robot can reach Region $4$ or $5$ from Region $2$ under the control action $c_1$, but it cannot decide for sure which region it will reach. We choose $\mathcal{AP}=\{ P_{1}, P_{2} \}$ and the labeling function is defined by $L(2)=\{P_{1}\}$, $L(6)=\{P_{2}\}$, and $L(s)=\emptyset$ otherwise.

The task of the robot is to reach Region $2$ first and then reach Region $6$ eventually, which can be expressed by the sc-LTL formula
\[
\varphi =\Diamond (P_{1} \wedge \Diamond P_{2}).
\]
The (modified) DFA satisfying  $\mathcal{L}(A_\varphi)=\textsf{Word}_{pref}(\varphi)$ is shown in Fig.~\ref{fig:moauto2}. The product system is presented in Fig.~\ref{fig:exm1}, where 
$X_{F}=\{ x_{6} \}$ and $X_{F \cup \{s_F \}}=\{x_{6},x_{7} \}$. Since both controller and intruder have full state information of the robot, we have $\tilde{H}((x,s))=H(x)=x$, i.e., output function is an identity map.

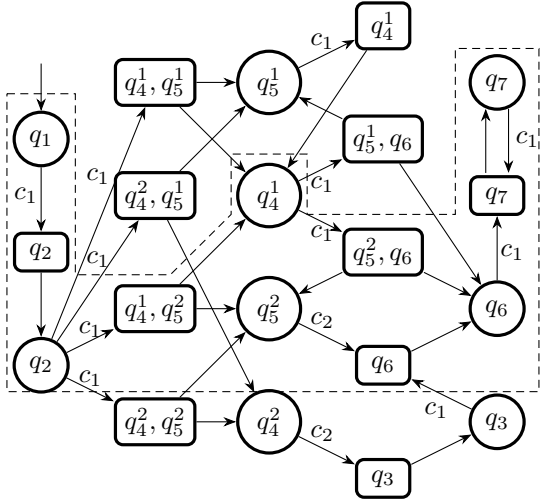
\begin{figure} 
  \centering
\begin{tikzpicture}
[
zstate/.style={rectangle, draw=black!255, fill=white!255, very thick, minimum height=5mm,minimum width=7mm, rounded corners = 1mm},
ystate/.style={circle, draw=black!255, fill=white!255, very thick, minimum height=5mm,minimum width=5mm},
]
    \node [ystate](q1)at(0,1.5){$q_{1}$};
	\node [zstate](q2)at(0,0){$q_{2}$};
	\node [ystate](q3)at(0,-1.5){$q_{2}$};
	\node [zstate](q4)at(1.5,2.25){$q_{4}^{1},q_{5}^{1}$};
	\node [zstate](q5)at(1.5,0.75){$q_{4}^{2},q_{5}^{1}$};
	\node [zstate](q6)at(1.5,-0.75){$q_{4}^{1},q_{5}^{2}$};
	\node [zstate](q7)at(1.5,-2.25){$q_{4}^{2},q_{5}^{2}$};
    \node [ystate](q8)at(3,2.25){$q_{5}^{1}$};
	\node [ystate](q9)at(3,0.75){$q_{4}^{1}$};
	\node [ystate](q10)at(3,-2.25){$q_{4}^{2}$};
	\node [ystate](q11)at(3,-0.75){$q_{5}^{2}$};
	\node [zstate](q12)at(4.5,3){$q_{4}^{1}$};
	\node [zstate](q13)at(4.5,1.5){$q_{5}^{1},q_{6}$};
	\node [zstate](q14)at(4.5,0){$q_{5}^{2},q_{6}$};
    \node [zstate](q15)at(4.5,-3){$q_{3}$};
	\node [zstate](q16)at(4.5,-1.5){$q_{6}$};
	\node [ystate](q17)at(6,2.25){$q_{7}$};
	\node [zstate](q18)at(6,0.75){$q_{7}$};
	\node [ystate](q19)at(6,-0.75){$q_{6}$};
	\node [ystate](q20)at(6,-2.25){$q_{3}$};

	\node (c12)at(-0.2,0.75){$c_{1}$};
	\node (c34)at(0.75,1){$c_{1}$};
	\node (c35)at(0.75,-0.1){$c_{1}$};
	\node (c36)at(0.65,-1.025){$c_{1}$};
	\node (c37)at(0.65,-1.675){$c_{1}$};
	\node (c812)at(3.7,2.8){$c_{1}$};
	\node (c913)at(3.7,0.9){$c_{1}$};
	\node (c914)at(3.7,0.25){$c_{1}$};
	\node (c1015)at(3.7,-2.4){$c_{2}$};
	\node (c1116)at(3.7,-0.9){$c_{2}$};
	\node (c1718)at(6.4,1.5){$c_{1}$};
	\node (c1918)at(6.2,0){$c_{1}$};
	\node (c2016)at(5.2,-2.1){$c_{1}$};

\draw [-{Stealth}](0,2.5) --(0,1.87);
\draw [-{Stealth}](5.85,1.05) --(5.85,1.9);
\draw [-{Stealth}](6.15,1.9) --(6.15,1.05);
\draw[-{Stealth}] (q1) -- (q2);
\draw[-{Stealth}] (q2) -- (q3);
\draw[-{Stealth}] (q3) -- (q4);
\draw[-{Stealth}] (q3) -- (q5);
\draw[-{Stealth}] (q3) -- (q6);
\draw[-{Stealth}] (q3) -- (q7);
\draw[-{Stealth}] (q4) -- (q8);
\draw[-{Stealth}] (q4) -- (q9);
\draw[-{Stealth}] (q5) -- (q8);
\draw[-{Stealth}] (q5) -- (q10);
\draw[-{Stealth}] (q6) -- (q9);
\draw[-{Stealth}] (q6) -- (q11);
\draw[-{Stealth}] (q7) -- (q10);
\draw[-{Stealth}] (q7) -- (q11);
\draw[-{Stealth}] (q8) -- (q12);
\draw[-{Stealth}] (q9) -- (q13);
\draw[-{Stealth}] (q9) -- (q14);
\draw[-{Stealth}] (q10) -- (q15);
\draw[-{Stealth}] (q11) -- (q16);
\draw[-{Stealth}] (q12) -- (q9);
\draw[-{Stealth}] (q13) -- (q19);
\draw[-{Stealth}] (q13) -- (q8);
\draw[-{Stealth}] (q14) -- (q19);
\draw[-{Stealth}] (q14) -- (q11);
\draw[-{Stealth}] (q15) -- (q20);
\draw[-{Stealth}] (q16) -- (q19);
\draw[-{Stealth}] (q19) -- (q18);
\draw[-{Stealth}] (q20) -- (q16);

\draw[densely dashed] (-0.45,2.1)--(0.45,2.1)--(0.45,-0.3)--(1.7,-0.3)--(2.5,0.5)--(2.5,1.3)--(3.5,1.3)--(3.5,0.5)--(5.45,0.5)--(5.45,2.7)--(6.55,2.7)--(6.55,-1.85)--(-0.45,-1.85)--cycle;

\end{tikzpicture} 
  \caption
    {$AES(G)$. Circle states are $Y$-states, rectangular states are $Z$-states. Predictions for each state are $h_{1}=(0,0,0,0)$,  $h_{2}=(0,0,0,0)$, $h_{3}=(0,0,0,0)$, $h_{4}^{1}=(0,0,0,0)$, $h_{4}^{2}=(0,0,1,0)$, $h_{5}^{1}=(0,0,0,0)$, $h_{5}^{2}=(0,1,0,0)$, $h_{6}=(1,0,0,0)$, $h_{7}=(0,0,0,0)$, where $h_i$ denotes the prediction for $q_i$ and $h_i^j$ for $q_i^j$. }
 \label{fig:AES}
\end{figure}

Let us consider parameter $K=3$, which means that the intruder should not predict the exact satisfication time of the robot $3$ steps ahead.
The $AES(G)$ is shown in Fig.~\ref{fig:AES}, where circle states represent $Y$-states and rectangular states represent $Z$-states. We use $q_i$ to denote the augmented state such that $\textsf{state}(q_{i})=x_{i}$. Also, for different $j$, $q_{i}^{j}$ represent different predictions for $x_{i}$. For instance, $\textsf{pred}(q_{4}^{1})=(0,0,0,0)$ and $\textsf{pred}(q_{4}^{2})=(0,0,1,0)$. The deterministic BTS $T$ in line 8 of Algorithm \ref{alg:security} is presented by the part within the dashed line of Fig.~\ref{fig:AES}. The controller $C$ decoded from it works as follows: $C(x_{1})=c_{1}$, $C(x_{1}x_{2})=c_{1}$, $C(x_{1}x_{2}x_{4})=c_{1}$, $C(x_{1}x_{2}x_{5})=c_{2}$, $C(x_{1}x_{2}x_{4}x_{5})=c_{2}$, $C(\tau)=c_{1}$ for any other $\tau \in \textsf{Path}(G_{C})$. Regarding predictions, for example, we have $h=\textsf{pred}(q_{4}^{1})=(0,0,0,0)$, $H=\textsf{pred}(\{ q_{5}^{2},q_{6} \})=\{ (0,1,0,0),(1,0,0,0) \}$, and $(h, H) \in \Delta$. 
The reasons are as followings: 
(i) $h[0]=0$ since  $x_{4} \notin X_{F}$; and 
(ii) $h[1]=0$ since $x_{4}$ has the successor $x_5$; and  (iii) for $i\in\{2,3\},$ $h[i]=0$ since $x_{4}x_{6}(x_{7})^{*}$ are paths starting from $x_{4}$ whose last elements are not in $X_{F}$.

Thus, this controller may generate three possible paths shown  
as the colored lines  in Fig.~\ref{fig:tra1}.
As we have discussed previously, all these three paths are $3$-step unpredictable.
Essentially, here even when the system is fully observed, we still could leverage the  uncertainty of the transition system to confuse intruder on when robot will reach the secret state, i.e., finish the task.

\section{Conclusion}
In this paper, we formulated and solved a security-aware controller synthesis problem. 
The synthesized controller can ensure both a given scLTL task and the unpredictability of the satisfaction time of the task. 
A novel information structure incorporating the effect of control decisions in the future  was provided.  
We show that our synthesis algorithm is both sound and complete to the problem. 
In this work, we assume that the controller and the intruder have the same observation. 
In the future, we plan to relax this assumption by considering controllers and intruders with incomparable observation. 

                                                   







\bibliography{ifacconf}

\newpage 

\appendix
\section{Proof} \label{ap:proof}
$\mathbf{Proof \text{ } of \text{ } Proposition}$~\ref{prop:trans}
The ``only if'' part is trivial and we prove the ``if'' part  by contradiction. Assume that $\tilde{G}_{C}$ is not $K$-step unpredictable, then there exists a path $\tau \in \textsf{Path}(\tilde{G}_{C})$ which is not $K$-step unpredictable, i.e., $\exists m \geq K$ such that for all $\tau_{1}\tau_{2} \in \textsf{Path}(\tilde{G}_{C})$ satisfying $\tilde{H}(\tau)=\tilde{H}(\tau_{1})$ and $|\tau_{2}|=m$, we have $\textsf{last}(\tau_{2}) \in \tilde{X}_{F}$. Consider $\tau'=\tau\tau'_1 \in \textsf{Path}(\tilde{G}_{C})$, where $|\tau'_1|=m-K$. Then for any $\tilde{\tau}_{1}\tilde{\tau}_{2} \in \textsf{Path}(\tilde{G}_{C})$ such that $\tilde{H}(\tau')=\tilde{H}(\tilde{\tau}_{1})$ and $|\tilde{\tau}_{2}|=K$, we have that $\textsf{last}(\tilde{\tau}_{2}) \in \tilde{X}_{F}$. This violates the condition of proposition. Thus ``if'' holds.

$\mathbf{Proof \text{ } of \text{ } Proposition}$~\ref{5-16}
We prove by induction on the length of paths in $G _{C}$.

Induction Basis: consider $\tau \in \textsf{Path}(G _{C})$ with $|\tau|=1$.

The unique path whose length is 1 in $G _{C}$ is the path that only contains initial state $x_{0}$. Then, $Y_{T}(H(x_{0}))=y_{0}$ is initial state of $T$. We have $\mathcal{E}_{C}(H(x_{0}))=\{ x_{0} \}=\textsf{state}(y_{0})$.

Induction Step:
assume that the induction hypothesis is true when the path length is $k$.

Consider $ \tau \in \textsf{Path}(G_{C})$ with $|\tau|=k$ and let $y=Y_{T}(H(\tau))$. Since $T$ is complete, we can find $z=\delta_{YZ}(y,U_{T}(y))$. For any $\hat{x},\hat{x}' \in \textsf{state}(y)$, we have $H(\hat{x})=H(\hat{x}')$. Since $\textsf{state}(z) \neq \emptyset$ and we assume that states with same output have same active control input set, we know that $\exists x \in X$ such that $(\textsf{last}(\tau),U_{T}(y),x) \in \rightarrow$. Therefore, every path in $G_{C}$ with length $k$ has successors. Moreover, since $T$ is complete, we know that for $x' \in X$ and $\tau'=\tau x' \in \textsf{Path}(G_{C})$, $Y_{T}(H(\tau'))=\delta_{ZY}(\delta_{YZ}(y,U_{T}(y)),H(x'))$. Therefore, $Y_{T}(H(\tau'))$ is well-defined.

Now, we consider path $\tau \in \textsf{Path}(G _{C})$ with $|\tau|=k+1$. Let $y=Y_{T}(H(\tau))$. Let $\tau=\tau' x$ such that $x \in X$ and $y'=Y_{T}(H(\tau'))$. From induction hypothesis, we have $\textsf{state}(y')=\mathcal{E}_{C}(H(\tau'))$. Then

$\quad \mathcal{E}_{C}(H(\tau)) $ 

$ =\{ x' \in X  | \tau'' \in \textsf{Path}(G_{C}) $, $ H(\tau'')=H(\tau) $, $ \textsf{last}(\tau'')=x' \}$

$ = \{ x' \in X  | \tau'' \in \textsf{Path}(G _{C}) $, $ H(\tau'')=H(\tau') $, $ c=U_{T}(y') , $

$\qquad \qquad \qquad \qquad \quad (\textsf{last}(\tau''),c,x') \in \rightarrow  $, $ H(x')=H(x) \} $

$\stackrel{(1)}{=} \{ x' \in X  | x'' \in \textsf{state}(y')$, $c=U_{T}(y') ,$

$\qquad \qquad \qquad \qquad  (x'',c,x') \in \rightarrow  $, $ H(x'')=H(x')  \}$

$\stackrel{(2)}{=} \textsf{state}(y) $.

$(1)$ is right because induction hypothesis is true for $k$. The correctness of $(2)$ comes from the definition of $Y$-$Z$ transition and $Z$-$Y$ transition of BTS. This completes the proof.

$\mathbf{Proof \text{ } of \text{ } Proposition}$~\ref{5-9}
From Proposition \ref{5-16}, for $\tau \in \textsf{Path}(G_{C})$ and $y=Y_{T}(H(\tau))$, we have $\textsf{state}(y)=\mathcal{E}_{C}(H(\tau))$. Thus, $\textsf{last}(\tau) \in \textsf{state}(y)$. Since $y \in \mathcal{B}$, we know that there exists unique $\hat{x}=(\textsf{last}(\tau),h) \in y$.

We prove the left part of this proposition by inducting the length of prediction horizon.

Induction Basis: let's consider the case of $i=0$.

We need to prove that $h(0)=1$, which is equivalent to $\textsf{Reach}^{0}_{G _{C}}(\tau) \subseteq X_{F}$. Obviously, we have $\textsf{Reach}^{0}_{G _{C}}(\tau)=\{\textsf{last}(\tau)\}$. Thus, $\textsf{Reach}^{0}_{G _{C}}(\tau) \subseteq X_{F} \Leftrightarrow \textsf{last}(\tau) \in X_{F}$. Since $y \in \hat{X}_{cons}$, $\textsf{last}(\tau) \in X_{F} \Leftrightarrow h(0)=1$. This completes the proof of induction basis.

Induction Step:

Assume that the induction hypothesis is true when $i=n < K$. 
Then, we have
\begin{equation}  \label{pf:1}
    \textsf{Reach}^{n+1}_{G _{C}}(\tau)=\bigcup_{\tau' \in X_{N}(\tau)} \textsf{Reach}^{n}_{G _{C}}(\tau'),
\end{equation}
where $X_{N}(\tau)=\{ \tau x \in \textsf{Path}(G _{C}) |  x \in X \}$ is the set of paths in $G _{C}$ that has one more state concatenated to $\tau$.

We also define 
\begin{align}
    Z_{N}(\tau)&= \{ \hat{x}' |\hat{x}' \in y' , y' = Y_{T}(H(\tau')),\nonumber\\
    &\quad\quad\quad\tau' \in X_{N}(\tau), \textsf{last}(\tau')=\textsf{state}(\hat{x}') \}
\end{align}
as the set of corresponding $Y$-states of paths in $X_{N}(\tau)$. Since in BTS, all $Y$-$Z$ transitions are one-step consistent, we have that
\begin{equation}\label{pf:2}
   h(n+1)=1 \Leftrightarrow \forall \hat{x}' \in  Z_{N}(\tau), \textsf{pred}(\hat{x}')(n)=1.
\end{equation}
 From induction hypothesis we know that 
 \begin{equation} 
    \textsf{pred}(\hat{x}')(n)=1 \Leftrightarrow \textsf{Reach}^{n}_{G _{C}}(\tau') \subseteq X_{F}, \nonumber
 \end{equation}
 where $\hat{x}' \in y'$, $y'=Y_{T}(H(\tau'))$, $\textsf{last}(\tau')=\textsf{state}(\hat{x}')$.

 Thus, we have:
 \begin{equation} \label{pf:3}
    \forall \hat{x}' \in  Z_{N}(\tau), \textsf{pred}(\hat{x}')(n)=1 \Leftrightarrow \bigcup_{\tau' \in X_{N}(\tau)} \textsf{Reach}^{n}_{G _{C}}(\tau') \subseteq X_{F}.
 \end{equation}
 Combine~(\ref{pf:1})-(\ref{pf:3}) and we will get:
 \[
 \forall \tau \in \textsf{Path}(G_{C}): \textsf{Reach}^{n+1}_{G _{C}}(\tau) \subseteq X_{F} \Leftrightarrow h(n+1)=1.
 \]
 This completes the proof.

$\mathbf{Proof \text{ } of \text{ } Proposition}$~\ref{5-10}
 The result of Proposition \ref{5-9} shows that for any $\tau \in \textsf{Path}(G_{C})$, $\textsf{Reach}^{K}_{G_{C}}(\tau) \nsubseteq X_{F} \Leftrightarrow \textsf{pred}(\hat{x})(K)=0$, where $\hat{x} \in y$ such that $\textsf{state}(\hat{x})=\textsf{last}(\tau)$ and $y=Y_{T}(H(\tau))$. From Proposition~\ref{5-16}, we know that for $\tau \in \textsf{Path}(G_{C})$, $\textsf{state}(y)=\mathcal{E}_{C}(H(\tau))$. From result of Proposition \ref{prop:trans}, $G_{C}$ is $K$-step unpredictable if and only if for path $\tau \in \textsf{Path}(G_{C})$, there exist $\tau' \in \textsf{Path}(G _{C})$ such that $H(\tau)=H(\tau')$ and $\textsf{Reach}^{K}_{G _{C}}(\tau') \nsubseteq X_{F}$, i.e., $\exists \hat{x} \in y$ s.t. $\textsf{pred}(\hat{x})(K)=0$. Therefore, if for any $\tau \in \textsf{Path}(G_{C})$, $Y_{T}(H(\tau)) \notin \mathcal{B}_{ins}$, $G_{C}$ is $K$-step unpredictable. It is true from definition of BTS. This completes the proof.

$\mathbf{Proof \text{ } of \text{ } Theorem}$~\ref{thm:souandcom}
We first prove that Algorithm~\ref{alg:security} is sound. Assume that $T$ is the deterministic BTS in line 8 of Algorithm~\ref{alg:security}.
From Proposition~\ref{5-10} and Proposition~\ref{5-16}, we know that $G_{C}$ is live and $K$-step unpredictable. If $Q_{Y}^{F(k)}\cup Q_{Z}^{F(k)} \neq Q_{Y}^{F(k+1)} \cup Q_{Z}^{F(k+1)}$, then $Q_{Y}^{F(m)}\cup Q_{Z}^{F(m)} \neq Q_{Y}^{F(m+1)} \cup Q_{Z}^{F(m+1)}$ for $0 \leq m \leq k$. Therefore, $Q_{Y}^{F}=\bigcup_{k=0}^{M}Q_{Y}^{F(k)}$, where $M=|Q_{Z}^{AES}|+|Q_{Y}^{AES}|$. For $\tau \in \textsf{Path}(G_{C})$, if $Y_{T}(H(\tau)) \in Q_{Y}^{F(k)}$, then $\textsf{last}(\tau') \in X_{F\cup \{ s_{F}\}}$ holds for any $\tau\tau' \in Path(G_{C})$ with $|\tau'|=k$. Therefore, if $|\tau| \geq M$, we know that $\textsf{last}(\tau) \in X_{F\cup \{ s_{F}\}}$. Thus, $G_{C} \models \varphi$ is true and Algorithm~\ref{alg:security} is sound.

Next, we prove the completeness of Algorithm~\ref{alg:security} by contradiction. Assume that there exists a live controller $C$ such that $G_{C} \models \varphi$ and $G_{C}$ is $K$-step unpredictable, and assume that Algorithm~\ref{alg:security} returns ``no solution''.

We first prove that $AES(G)$ is not empty. Towards this end, we define a function $U: \textsf{Path}(G_{C}) \rightarrow \hat{X} $, where for $\tau \in \textsf{Path}(G_{C})$ and $0 \leq i \leq K$, we have $\textsf{state}(U(\tau))=\textsf{last}(\tau)$, and $\textsf{pred}(U(\tau))(i)=1$, which is equivalent to $R_{G_{C}}^{i}(\tau) \subseteq X_{F}$.

We then define
\[
P=\{ o \in O^{*} |\tau \in \textsf{Path}(G_{C}), o=H(\tau) \}
\]
as the set of observations of paths in $G_{C}$. Then for $o \in P$, we define 
\[
S_{Y}(o)=\{ U(\tau) | \tau \in \textsf{Path}(G_{C}) , H(\tau)=o \}.
\]
Also, for $o \in P$ and $c = C(o)$, we define
\[
S_{Z}(o,c)=\{ U(\tau x) | \tau x \in \textsf{Path}(G_{C}),  x \in X , 
\]
\[
\qquad \qquad \qquad \qquad H(\tau)=o , (\textsf{last}(\tau),c,x) \in \rightarrow \}.
\]
We let
\[
S_{Y} = \{ S_{Y}(o) \subseteq \hat{X} | o \in P \},
\]
and
\[
S_{Z} = \{ S_{Z}(o,c) \subseteq \hat{X} | o \in P , c = C(o)  \}.
\]
Then, we define $S_{YZ}: S_{Y} \times U \rightarrow 2^{T_{Z}}$ by
\[
S_{YZ}(y,c) = \{ S_{Z}(o,c) \in S_{Z} | o \in P , y=S_{Y}(o), c=C(o) \},
\]
and $S_{ZY}: S_{Z} \times O \rightarrow S_{Y}$ by
\[
S_{ZY}(z,o)= \{ x \in z | H(x)=o \}.
\]
Let $y_{0}=U(x_{0})$. We now prove that $(S_{Y},S_{Z},S_{YZ},S_{ZY},y_{0})$ is the subsystem of $AES$. Since $G_{C}$ is live, we know that for any $y \in T_{Y}$, there exist $c \in U$ and $z \in S_{Z}$ such that $z \in S_{YZ}(y,c)$, which means that $Y$-states in $S_{Y}$ are complete. Then, based on the definition of $S_{ZY}$, we know that all $Z$-states in $S_{Z}$ are also complete. Since $G_{C}$ is $K$-step unpredictable, we have that $y \in \mathcal{B}_{sec}$ holds for any $y \in S_{Y} $. For any $\tau \in \textsf{Path}(G_{C})$, let $x = U(\tau)$. Since $\textsf{pred}(x)(i)$ represents whether or not robot will definitely reach states in $X_{F}$ in exact $i$ steps when following path $\tau$, all transitions defined by $S_{YZ}$ are one-step consistent. It means that $(S_{Y},S_{Z},S_{YZ},S_{ZY})$ satisfies conditions of $AES$. Thus $AES$ is not empty.

Since $C$ satisfies task completion requirement, we know that $S_{Y} \subseteq Q_{Y}^{F}$. Otherwise, there exists a path in $G_{C}$ which never reaches $X_{F \cup \{ s_{F} \}}$. Also, $y_{0}^{AES}\cap S_{Y} \neq \emptyset$ holds. Thus, there exists a solution to Algorithm~\ref{alg:security}, which contradicts with the assumption that Algorithm~\ref{alg:security} returns ``no solution''. This implies the completeness of Algorithm~\ref{alg:security}.

\end{document}